# Approximated Computation of Belief Functions for Robust Design Optimization

Massimiliano Vasile[1] and Edmondo MInisci[2]
*University of Strathclyde, G1 1XJ , Glasgow, UK*

Quirien Wijnands[3]
*ESA/ESTEC, 2200 AG, Noordwijk, The Netherlands*

This paper presents some ideas to reduce the computational cost of evidence-based robust design optimization. Evidence Theory crystallizes both the aleatory and epistemic uncertainties in the design parameters, providing two quantitative measures, Belief and Plausibility, of the credibility of the computed value of the design budgets. The paper proposes some techniques to compute an approximation of Belief and Plausibility at a cost that is a fraction of the one required for an accurate calculation of the two values. Some simple test cases will show how the proposed techniques scale with the dimension of the problem. Finally a simple example of spacecraft system design is presented.

## Nomenclature

| | | |
|---|---|---|
| $\alpha_{SA}$ | = | solar array solar aspect angle, rad |
| $\Delta t_a$ | = | access time, s |
| $\Delta t_{aq}$ | = | acquisition time, s |
| $\rho_{SA}$ | = | solar array specific mass, kg/m$^2$ |
| $\rho_{CMR}$ | = | amplifier case mass fraction |
| $\rho_A$ | = | antenna specific mass, kg/m$^2$ |
| $\eta_{cell}$ | = | solar cell efficiency |
| $\eta_b$ | = | battery efficiency |
| $\eta_{PCU}$ | = | PCU efficiency |
| $\nu$ | = | generic threshold |
| $\theta$ | = | focal element |
| $\varphi_f$ | = | Faraday rotation, rad |
| $Pl$ | = | cumulative Plausibility function |
| $A$ | = | generic proposition |
| $A_\nu$ | = | archive |
| $A_L$ | = | atmospheric losses, dB |
| $AM_L$ | = | antenna misalignment loss, dB |
| $AN_{Temp}$ | = | antenna noise temperature, K |
| $a_{PCU}$ | = | PCU mass coefficient, kg/W |
| $B$ | = | onboard data volume, bits |
| $Bel$ | = | cumulative Belief function |
| $B_{l,i}$ | = | generic box in $\bar{U}$ |
| $C_{min}$ | = | battery capacity, Wh |
| $c$ | = | speed of light, m/s |
| $D_{ant}$ | = | antenna diameter, m |
| $D$ | = | design space |
| $DOD$ | = | Depth Of Discharge |
| **d** | = | design parameter vector |

---

| | | |
|---|---|---|
| $E_d$ | = | battery energy density, Wh/kg |
| $e$ | = | elevation error, rad |
| $F_L$ | = | feeder loss, dB |
| $FS_L$ | = | free space loss, dB |
| $f,g,h$ | = | generic functions |
| $f_T$ | = | carrier frequency, MHz |
| $G_{AMP}$ | = | amplifier gain, dB |
| $G_r$ | = | ground station gain, dB |
| $H_G$ | = | ground station altitude, m |
| $I_d$ | = | inherent degradation |
| $I_L$ | = | implementation loss, dB |
| $L_d$ | = | time degradation |
| $P_0$ | = | generated power per unit area, W/m$^2$ |
| $P_d$ | = | power in daylight, W |
| $P_L$ | = | polarization mismatch loss, dB |
| $P_{Ld}$ | = | required transmission power, W |
| $P_e$ | = | power in eclipse, W |
| $P_{SA}$ | = | total required power, W |
| $R_t$ | = | data rate, dB |
| $RA$ | = | rain absorption |
| $R_{aL}$ | = | rain absorption loss, dB |
| $r_{GS}$ | = | distance from the ground station, km |
| $S_{LAT}$ | = | horn lateral surface, m$^2$ |
| $T$ | = | amplifier type |
| $T_{AMP}$ | = | amplifier noise, K |
| $T_{data}$ | = | transmitted data, bits |
| $U$ | = | uncertain space |
| **u** | = | uncertain parameter vector |
| $X_e$ | = | power system efficiency in eclipse |
| $X_d$ | = | power system efficiency in daylight |

## I. Introduction

IN recent times, Evidence Theory has been proposed in place of Probability Theory for robust design of engineering systems. Authors like Oberkampf et al.[1] demonstrated the potentiality of Evidence Theory to model both epistemic and aleatory uncertainties in the design of engineering systems. Similar examples can be found in the work of Agarwal et al.[2] or in the work of Bae et al.[3], Fetz et al.[4] He et al.[5] and Mourela et al.[6], mainly with applications to structural design. Denoeux proposed a technique to compute an inner and outer approximation of Belief and Plausibility functions[7]. Helton et al.[8,9,10] proposed a number of techniques to reduce the dimensionality of problems treated with Evidence-based models. More recently Vasile[11] and Croissard et al.[12] provided some examples of application of Evidence Theory to the robust optimal design of space systems and space trajectories. The uncertainties in the design parameters of the main spacecraft subsystems were modeled using Evidence Theory. The design process was then formulated as an Optimization Under Uncertainties (OUU) and the Belief function was optimized (maximized) together with all the other criteria that define the optimality of the system.

With Evidence Theory, also know as Dempster-Shafer's theory[13], both aleatory and epistemic uncertainties, coming from a poor or incomplete knowledge of the design parameters, can be correctly modeled. The values of uncertain or vague design parameters can be expressed by means of intervals with associated basic belief assignment or *bpa*. Each expert participating in the design, assigns an interval and a *bpa* according to their experience. Ultimately, all the pieces of information associated to each interval are fused together to yield two cumulative values, Belief and Plausibility, that express the confidence range in the optimal design point. In particular, the value of Belief expresses the lower limit on the probability that the selected design point remains optimal (and feasible) even under uncertainties. The benefits coming from the use of Evidence Theory are considerable but the computation of Belief and Plausibility requires running a number of optimizations that grows exponentially with the number of dimensions and becomes intractable even for problems of moderate size.

This paper presents some ideas on how to reduce the computational cost to obtain an approximation of Belief and Plausibility cumulative functions in space system engineering. Some of the techniques presented in this paper are not problem dependent others exploit the partial decomposability of space system engineering design problems. The paper starts with a brief introduction to Evidence Theory and its use in the context of robust design optimization. It then presents some techniques to compute an optimal design solution under uncertainty when Evidence Theory is used for uncertainty quantification. A few ideas are then proposed to reduce the computational cost and their effectiveness is experimentally proven on some scalable analytic functions. The preliminary robust design of an integrated power and telecommunication system of a satellite is then used to illustrate the application of Evidence-based Robust Design Optimization to the design and margin quantification of space systems. A final section introduces a problem decomposition technique that looks promising to solve large scale space system engineering problems in polynomial time.

## II  EVIDENCE-BASED ROBUST DESIGN OPTIMISATION

Evidence Theory, developed by Shafer[13], belongs to the class of imprecise probability theories conceived to adequately treat both epistemic and aleatory uncertainty when no information of probability distributions is available. The theory does not require additional assumptions when the available information is poor or incomplete and provides a nice framework to incorporate multiple pieces of evidence in support to a statement. In most current engineering design applications of Evidence Theory, domain experts are expected to express their belief on the value of an uncertain parameter *u* being within a certain set of intervals. Each interval can be considered as an elementary proposition, and all the intervals form the so-called frame of discernment $\Theta$, which is a set of mutually exclusive elementary propositions. The frame of discernment can be viewed as the counterpart of the finite sample space in probability theory. The power set of $\Theta$ is $U=2^{\Theta}$ or the set of all the subsets of $\Theta$ (the uncertain space in the following). The level of confidence an expert has in an element $\theta$ of $U$ is quantified using the Basic Probability Assignment (*bpa*) $m(\theta)$ that satisfies the axioms:

$$m(\theta) \geq 0, \forall \theta \in U = 2^{\Theta};$$
$$m(\theta) = 0, \forall \theta \notin U = 2^{\Theta};$$
$$m(\varnothing) = 0;$$
$$\sum_{\theta \in U} m(\theta) = 1$$
(1)

Note that the *bpa* is actually a belief in the values of $\theta$ rather than an actual probability. An element of $U$ that has a non-zero *bpa* is named a focal element $\theta$. When more than one parameter is uncertain, the focal elements are the result of the Cartesian product of all the intervals associated to each uncertain parameter. The *bpa* of a given focal element is then the product of the *bpa* of each interval. All the pieces of evidence completely in support of a given proposition form the cumulative belief function *Bel* while all the pieces of evidence partially in support of a given proposition from the cumulative plausibility function *Pl*. In mathematical terms the two functions are defined as follows:

$$Bel(A) = \sum_{\forall \theta_i \subseteq A} m(\theta_i)$$
(2)

$$Pl(A) = \sum_{\forall \theta_i \cap A \neq 0} m(\theta_i)$$

where *A* is the proposition about which the Belief and Plausibility need to be evaluated. For example, the proposition can be expressed as:

$$A = \{\mathbf{u} \in U \mid f(\mathbf{u}) \leq \nu\}$$
(3)

where *f* is the system process and the threshold $\nu$ is the value of a design budget (e.g. the mass). Thus, focal elements intercepting the set *A* but not included in *A* are considered in *Pl* but not in *Bel*. It is important to note that the set *A* can be disconnected or present holes, likewise the focal elements can be disconnected or partially overlapping.

### A. Robust Design Formulation

The interest is in a general function $f: D \times U \subseteq \mathbb{R}^{m+n} \to \mathbb{R}$ characterizing an engineering system to be optimized, where *D* is here called the available design space and *U* the uncertain space. The function *f* represents the model of the system budgets (e.g. power budget, mass budget, etc.), and depends on some uncertain parameters **u** and design parameters **d** such that:

$$\mathbf{u} \in U \subseteq \mathbb{R}^n; \quad \mathbf{d} \in D \subseteq \mathbb{R}^m \quad (4)$$

A *bpa* is associated to the frame of discernment *U* of the uncertain parameters **u**. From the definition of *Bel* and *Pl* and from Eq. (3) it is clear that the maximum and minimum of *f* over every focal element of *U* should be computed and compared to ν. The threshold ν is the desired or expected value of the system budget. If the maximum and minimum do not occur at one of the vertices of the focal element an optimization problem has to be solved for every focal element and for each new design vector. Because the number of focal elements increases exponentially with the number of uncertain parameters and associated intervals so does the number of optimization problems. Furthermore, what designers are usually interested in are: a design solution **d** that optimizes performance (i.e. the design budgets) and minimizes the impact of uncertainty, a quantification of the design margins on the system budgets and a quantification of the reliability of the design solution. This information can be obtained for the worst case scenario but that might lead to over conservative decisions. Therefore it is desirable to have also the variation of the design margins and reliability with the threshold ν, i.e. with the expected value of the design budgets Indeed, it may be relevant to take a little more risk (a slightly lower value of the belief) if the performance gain is significant. Therefore, in practice, it would be desirable to have the trade-off curve, solution of the bi-objective optimization problem:

$$\max_{\mathbf{d} \in D \land \mathbf{u} \in U} Bel\left(f(\mathbf{d},\mathbf{u},) < \nu\right)$$
$$\min \nu \quad (5)$$

In previous works[7,7], the bi-objective problem (5) was approached directly with a multi-objective evolutionary optimizer working on the **d** and ν. The whole curve could be reconstructed with a population of agents converging to the optimal pairs of values [*Bel* ν]. However, the computational cost was driven by the identification of *A* and the number of focal elements included in it. The assumption was that the maxima and minima of *f* were occurring only at the corners of the focal element. The evaluation of the corners is in itself an operation that grows exponentially with the number of dimensions and is, anyway, not applicable to a general case.

In this paper we propose a different way of approaching the problem. First of all, the computation of the Belief function is performed by exploiting the following relationship:

$$Bel(A) = 1 - Pl(\neg A) \quad (6)$$

According to (2), the calculation of $Pl(\neg A)$ is computationally cheaper than the calculation of *Bel*(*A*). In fact, any subset of *U* that contains at least one value (even a single sample) above the threshold ν contributes to $Pl(\neg A)$. The computation of *Bel*(*A*) instead requires that all the elements of *A* are below the threshold.

## III COMPUTATIONAL APPROACH

Problem (5) would require the solution of a number of optimization problems that is exponentially increasing with the number of focal elements. However, if one is interested only in the maximization of the Belief and in the *f*, the exponential complexity can be avoided by solving the following two distinct problems over the Cartesian product of the unit hypercube $\bar{U}$ and *D*:

$$\nu_{\max} = \min_D \max_{\bar{U}} f(\mathbf{d},\mathbf{u}) \quad (7)$$

$$\nu_{\min} = \min_D \min_{\bar{U}} f(\mathbf{d},\mathbf{u}) \quad (8)$$

where $\bar{U}$ is the normalized collection of all the focal elements in *U*. In other words, all the focal elements in *U* are normalized with respect to the maximum range of the uncertain parameters and collected into a compact unit hypercube in which all the focal elements are adjacent and not overlapping. A point in the unit hypercube $\bar{U}$ is then mapped into the normal space *U* through the simple affine transformation:

$$\mathbf{x}_{U,i} = \frac{\left(\mathbf{b}_{U,i}^u - \mathbf{b}_{U,i}^l\right)}{\left(\mathbf{b}_{\bar{U},i}^u - \mathbf{b}_{\bar{U},i}^l\right)} \mathbf{x}_{\bar{U},i} + \mathbf{b}_{U,i}^l - \frac{\left(\mathbf{b}_{U,i}^u - \mathbf{b}_{U,i}^l\right)}{\left(\mathbf{b}_{\bar{U},i}^u - \mathbf{b}_{\bar{U},i}^l\right)} \mathbf{b}_{\bar{U},i}^l \quad (9)$$

where $\mathbf{b}^u_{\bar{U},i}$ and $\mathbf{b}^l_{\bar{U},i}$ are the upper and lower boundaries of the *i*-th hypercube to which $\mathbf{x}_{\bar{U},i}$ belongs and $\mathbf{b}^u_{U,i}$ and $\mathbf{b}^l_{U,i}$ are the upper and lower boundaries of the *i*-th hypercube to which $\mathbf{x}_{U,i}$ belongs. The transformation is relatively fast as it requires to scan only over the number of intervals per coordinate and not over the focal elements. The computational complexity of the affine transformation is, therefore, linear with the number of dimensions. The advantage is that each point within $\bar{U}$ belongs to at least one focal element, therefore by sampling $\bar{U}$ one is guaranteed to sample only the focal elements and not other parts of *U*.

Problem (7) looks for the minimum possible threshold value $\nu_{max}$ such that the entire unit hypercube is admissible, hence the Belief is 1. The solution of problem (7) does not require the exploration or even the generation of the focal elements and sets an upper limit on the value of the cost function. Problem (8) looks for the minimum threshold value $\nu_{min}$ above which the Plausibility is different from 0. As for problem (7), problem (8) does not require the knowledge of the focal elements and sets a lower limit on the value of the cost function. Below that limit the design is not feasible, given the current model and evidence on the design parameters.

The min/max problem is solved with a nested evolutionary process: an outer loop minimizes *f* over *D* and the inner loop maximizes *f* over $\bar{U}$. For each $\mathbf{d}_i$ vector an evolutionary process over $\bar{U}$ is run and the **u** vector with maximum *f* is associated to $\mathbf{d}_i$. The outer loop then proceeds till a maximum number of function evaluations is reached. For the inner loop a Matlab implementation of Inflationary Differential Evolution (IDEA) is used in this paper[15]. For the outer loop a modified version of IDEA, called vIDEA, is used. The modification is mainly in the way the objective function is computed. Due to the stochastic nature of the inner loop and the possible presence of multiple maxima the outcome of each inner loop might not be the global maximum or not even near to it. In order to increase the chances to produce optimal results, the local maximum $\mathbf{u}_{i,max}$, with objective value $f_{max}(\mathbf{d}_i,\mathbf{u}_{i,max})$, computed for $\mathbf{d}_i$ is compared against the local maximum $\mathbf{u}_{j,max}$, associated to design vector $\mathbf{d}_j$. If $f_{max}(\mathbf{d}_i,\mathbf{u}_{i,max}) < f_{max}(\mathbf{d}_i,\mathbf{u}_{j,max})$ then $\mathbf{u}_{j,max}$ and its related maximum are associated to $\mathbf{d}_i$. The underlying assumption here is that the cost function is Lipschitz continuous. Furthermore, if the location of the maxima in $\bar{U}$ does not change with **d** a full optimization for each **d** is not required. If instead the location of the maxima is changing with **d**, then for every $\mathbf{d}_i$ either a local search or a complete optimization is started. In both cases, running a full optimization or a simple local search depends on the vector difference between $\mathbf{d}_i$ at step *k* and at step *k*+1 of the evolution. The probability of running a full optimization is $P \sim p_d \|\mathbf{d}_i^{k+1} - \mathbf{d}_i^k\|$. The assumption here is that for small variations of **d** there are small variations of the location of the local maxima. This assumption is generally verified in the real-life applications the authors have encountered so far. A further level of verification of the global maximum is introduced $n_{pop}$ inner loop calls with probability $p_d$ by running a full optimization over $\bar{U}$ with two times the number of functions evaluations. IDEA implements a memetic type of evolutionary process in which a local search is started when the population collapses to a small region of the search space. The population is then restarted after the local search is completed and all the local minima are collected in an archive (for more details on the general algorithm implemented in IDEA the interested reader can refer to Ref. 15). In the inner loop, the local search is performed with a Quasi-Newton method and with a convergent Nelder-Mead approach in the outer loop. When the Nelder-Mead algorithm calls the inner loop no global search is run. At every restart of the population the archive is examined and the local minima are compared. As for the population, even for the archive the local maximum $\mathbf{u}_{i,max}$, with objective value $f_{min/max}(\mathbf{d}_{i,min},\mathbf{u}_{i,max})$, computed for $\mathbf{d}_{i,min}$ is compared against the local maximum $\mathbf{u}_{j,max}$, associated to local minimum design vector $\mathbf{d}_{j,min}$. If $f_{min/max}(\mathbf{d}_{i,min},\mathbf{u}_{i,max}) < f_{min/max}(\mathbf{d}_{i,min},\mathbf{u}_{j,max})$ then $\mathbf{u}_{j,max}$ and its related maximum are associated to $\mathbf{d}_{i,min}$. When a given number of function evaluations for the inner and the outer loop is exceeded, the search terminates. The best individual of the final population is added to the archive, the archive is ranked and the best values is validated running a final global optimization over $\bar{U}$ Because, the maximization and the minimization are based on stochastic processes a global convergence is not guaranteed unless the optimizer is globally convergent. Nonetheless, by using an evolutionary process the computational complexity remains polynomial. The consequence of an incorrect estimation of the global maximum and global minimum is an overestimation of the point with minimum Plausibility and an underestimation of the point with maximum Belief. However, on could argue that if a maximum is difficult to be found it correspond to a very unlikely event that legitimately provide little support to the evidence of a given proposition. Evolutionary process and uncertainty quantification are therefore closely entangled as the probability of sampling a particular uncertain value is directly related to the probability of the realization of the event that corresponds to that uncertain value.

### B. Evolutionary Binary Tree (EBT)

Given a design vector **d**, the value of *Bel* and *Pl*, for any ν value within [$\nu_{min}$ $\nu_{max}$], can be computed by building a binary tree in which a branch is pruned if the *max* of *f* associated to a leave is below (above) a given threshold $\bar{\nu}$. The binary tree is built as follows. The transformed uncertain space is partitioned by cutting every box $B_{l,i}$ (with $B_{0,0} = \bar{U}$) in two halves along the longest edge. The cutting point is the boundary of the uncertain interval that is the closest to the middle point of the edge. The tree is structured in levels with index *l* and at each level the tree has a number of leaves each one identified by the index *i*. Now let $B_{l,i}^l$ and $B_{l,i}^r$ be respectively the left and right halves deriving from the partition of the *i*-th box (or leave) $B_{l,i}$ at level *l*. If one box contains a maximum below the threshold or a minimum above the threshold the box is removed from the tree, otherwise it is added to the list of the boxes (leaves) that need to be partitioned at the following iteration (see Figure 1). If no maximum or minimum is yet computed either for $B_{l,i}^l$ or $B_{l,i}^r$ then the box is added to the list of those that need to be explored. The boxes that need to be explored and partitioned are said to be *undecidable* as a decision cannot be made on whether they contribute to the Belief or not. The exploration of an undecidable box is performed by running a global maximization and a global minimization of *f*(**d**,**u**) for a fixed **d** and for $\mathbf{u} \in B_{l,i}^l$ ($\mathbf{u} \in B_{l,i}^r$ respectively). In this implementation, IDEA is used for both the global maximization and minimization.

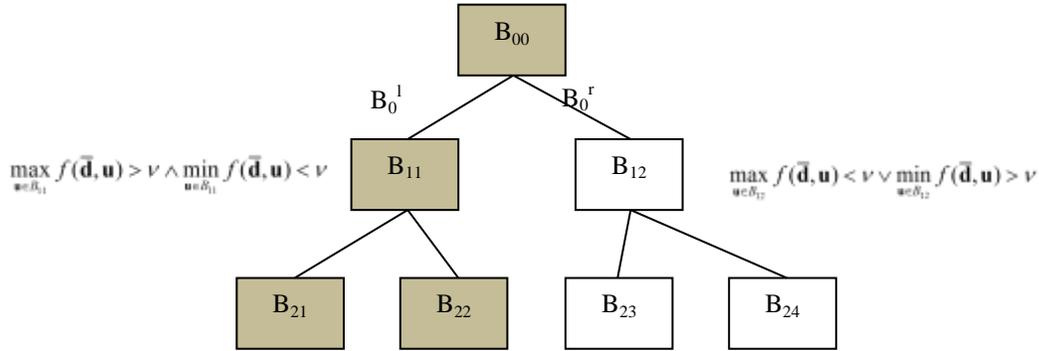

**Figure 1. EBRO Binary Tree**

Figure 1 shows a simple binary tree in which the right box deriving from splitting the initial uncertain space contains a maximum below the threshold or a minimum above the threshold. The whole branch of the tree is then discarded. The left box instead is undecidable and the branch that descends from that box needs to be explored. The branching and exploration proceed until a decision can be made for all the boxes. The exploration of a branch generates smaller and smaller boxes that eventually coincide with the focal elements. The exploration and branching process generates a number of focal elements clustered in macro boxes and a number of boxes that correspond to individual focal elements. All the discarded boxes with minimum above the threshold and all the boxes with maximum above the threshold are used to compute $Pl(\neg A)$. All the boxes with a minimum below the threshold are used to compute $Pl(A)$. The interesting aspect of this procedure is that $Pl(\neg A)$ and $Pl(A)$ can be computed even if all the maxima and minima are not identified exactly. In fact for a box to be included in the calculation of $Pl(\neg A)$ it is enough that a single value within the box, even not the actual maximum, is above the threshold. Likewise the calculation of $Pl(A)$ requires that even s single value, not necessary the actual minimum, is below the threshold.

Note that the collection of all the boxes generated for a given threshold $\bar{\nu}$ can be used to compute the Belief (Plausibility) values in the interval [$\bar{\nu}$ $\nu_{max}$]. In fact, the min and max values computed for each box represent additional ν values within [$\bar{\nu}$ $\nu_{max}$]. However, while the value of *Bel* and *Pl* can be exactly computed for $\bar{\nu}$ with the selected set of boxes, any other value within [$\bar{\nu}$ $\nu_{max}$] might result to be an underestimation of *Bel* and overestimation of *Pl* because of a lack of resolution (i.e. each box includes an excessive number of focal elements).

Therefore, if the whole curve is required, a refinement process is run by iteratively applying the evolutionary binary tree to the boxes that contain a given $v \in [\bar{v}\ v_{max}]$. This refinement process provides an exact value for each $v$ but can lead to a number of optimizations that is equal to two times the number of focal elements, if the entire curve is required. It is, however, interesting to note that if one takes an arbitrary $\bar{v} \in \left( v_{min}\ \frac{v_{max}+v_{min}}{2} \right)$, the boxes generated by the EBT can provide an approximation to the *Bel* and *Pl* curves that tend to be good in a neighborhood of $\bar{v}$ and close to $v_{max}$. This observation allows for the generation of a good estimation of the two curves at a fraction of the computational cost. Further to this approximation, two specific mechanisms have been devised to reduce the computational complexity and compute approximated *Bel* and *Pl* curves: the *integrated focal element filtering* and the *approximated min and max evaluation*.

*1. Integrated Focal Element filtering*

The *bpa* associated to each focal element decreases in magnitude as the number of dimensions increases. This observation suggests that an approximation to the value of *Bel* and *Pl* can be computed by using only a selected subset of focal elements. The reduction in *Bel* or increase in *Pl* due to this approximation can be quantified by looking at the cumulative value of the discarded focal elements. Therefore, during the generation of the binary tree boxes with a *bpa* below a given filter threshold are discarded until the cumulative *bpa* associated to those boxes is below the required filter accuracy

*2. Approximated Min and Max Evaluation*

Because the calculation of $Pl(\neg A)$ and $Pl(A)$ does not always require the exam maximum and minimum, one can consider using an approximation of the *min* and *max* values for each subdomain $B_{l,i}$ to make a decision. The evolutionary search for a maximum and a minimum proceed through an optimal sampling of the uncertain space. A selected subset of all the samples taken during the global exploration can be saved into an archive $A_v^u$, when maximizing, and into an archive into $A_v^l$ when minimizing, Then when two new subdomains $B_{l,i}^l$ and $B_{l,i}^r$ are generated by bisection of $B_{l,i}$, one can take the maximum element $\mathbf{x}_{sup}^l$ in $A_v^u \subseteq B_{l,i}^l$ (respectively $A_v^u \subseteq B_{l,i}^r$) and the minimum element $\mathbf{x}_{inf}^l$ in $A_v^l \subseteq B_{l,i}^l$ (respectively $A_v^l \subseteq B_{l,i}^r$) instead of running a full optimization for every new box.

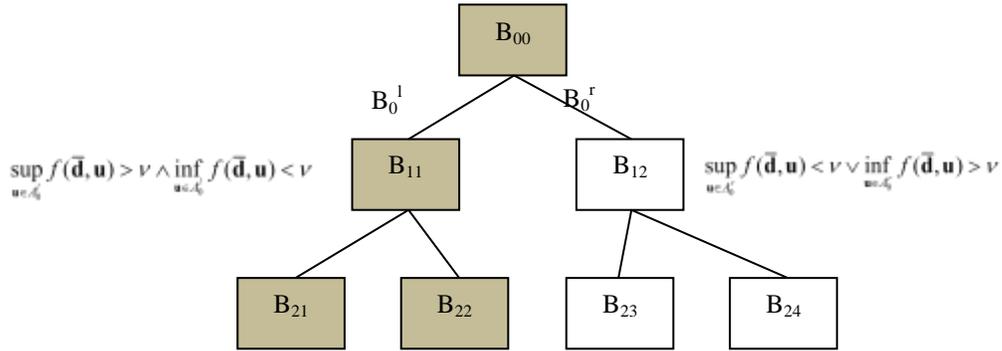

**Figure 2. Binary Tree with Approximated Min-Max Evaluation.**

Figure 2 shows a modified version of the binary tree in which decisions are made using the values in the archive instead of the actual maxima and minima. This technique produces a substantial reduction in the number of optimizations loops required to build the Belief and Plausibility curves but can introduce a significant error depending on the archiving procedure. In order to preserve accuracy the values $\mathbf{x}_{inf}^l, \mathbf{x}_{sup}^l, \mathbf{x}_{inf}^r$ and $\mathbf{x}_{sup}^r$ are trusted to be representative of the minimum and maximum values contained in the box with a probability $p_{trust}$ inversely proportional to the *bpa* associated to the box, i.e. the higher the *bpa* the lower the trust in the values $\mathbf{x}_{inf}^l$ and $\mathbf{x}_{sup}^l$ (respectively $\mathbf{x}_{inf}^r$ and $\mathbf{x}_{sup}^r$). In this way boxes with high *bpa* are properly explored with high probability. In particular, $\mathbf{x}_{\text{sup}}^{l}$ and $\mathbf{x}_{\text{sup}}^{r}$ are trusted if $p_{\text{trust}}>(\tau_c+bpa)$, where $\tau_c$ is a trust factor. The use of a this trust factor ensure good accuracy for a single value of $\nu$. During the refinement process, however, the EBT makes use of existing boxes but with different thresholds. Boxes that are correctly decided, on the basis of suboptimal archive values, for one threshold might not be correctly decided for a different threshold. When a new threshold is considered, the boxes generated for previous thresholds for which maximum and minimum was not exactly identified, are ranked from the one with highest *bpa* value to the one with the lowest. A maximization is then progressively run on each of them, starting from the one with highest *bpa*, until the cumulative *bpa* of the remaining ones is above $\tau_c$.

## IV TEST CASES

The effectiveness of the techniques to reduce the computational cost has been put to the test on a set of simple but representative, scalable problems (see Table 1). All the problems present a number of maxima and minima that grows with the number of dimensions. Problem MV1 for example has a number of maxima in $U$ that grows as $2^n$. Problem MV2 has maxima that change location with **d** while MV8 is multimodal and has the maxima that change with **d**. Table 2 represents the *bpa* structure for all the problems, with the intervals for each uncertain parameter and the associated *bpa*. Tests are run considering two possible different *bpa* structures: with a uniform distribution of *bpa* (labeled as EQ) and one with a non-uniform distribution. The tests in this paper are limited to three disconnected uncertain intervals per parameter but the results can be generalized to a higher number of intervals and also to overlapping intervals.

**Table 1. Standard Benchmark.**

| ID | Function | Parameters |
|---|---|---|
| MV1 | $f = \sum_{i=1}^{n} d_i u_i^2$ | $\mathbf{d} \in [1,5]^n$; $\mathbf{u} \in [-5,3]^n$ |
| MV2 | $f = \sum_{i=1}^{n} (u_i - d_i)^2$ | $\mathbf{d} \in [1,5]^n$; $\mathbf{u} \in [-5,3]^n$ |
| MV8 | $f = \sum_{i=1}^{n} (2\pi - u_i)\cos(u_i - d_i)$ | $\mathbf{d} \in [0,3]^n$; $\mathbf{u} \in [0,2\pi]^n$ |

**Table 2. BBA structure for the standard benchmark**

| MV1, MV2 | Interval | [-5 -4] | [-3 0] | [1 3] |
|---|---|---|---|---|
|  | bpa | 0.1 | 0.25 | 0.65 |
| MV8 | Interval | [0 1] | [2 4] | [5 2π] |
|  | bpa | 0.1 | 0.25 | 0.65 |
| MV1EQ,MV2EQ | Interval | [-5 -4] | [-3 0] | [1 3] |
|  | bpa | 0.33 | 0.33 | 0.34 |
| MV8EQ | Interval | [0 1] | [2 4] | [5 2π] |
|  | bpa | 0.33 | 0.33 | 0.34 |

Figure 3 presents the number of optimizations required to approximate the *Bel* and *Pl* curves for problem MV1 as a function of the number of dimensions, while Figure 4 presents the same result but for function MV2. Figure 5 presents the number of optimizations for problem MV8. Different trust factors and for a combination of the *integrated focal element filtering*, with a filter accuracy of 0.1, and the *approximated min and max evaluation*. The use of the *approximated min and max evaluation*, even with a trust factor of 0.99, leads to a reduction of the number of optimizations down to 25-30% of the number of optimizations required to explore all the focal elements. Reducing the trust factor to 0.90 leads to a further reduction and the combination with the integrated focal element filtering improves the reduction by 5-10%. The reduction is more limited in the case of a uniform distribution of *bpa*'s. Figure 6 shows the *Pl* and *Bel* for problem MV1 with n=6. This problem has a *Bel* curve with a steep drop at a thresh value of about 50. The figure shows the approximated curves using the *approximated min and max evaluation* and the combination *approximated min and max evaluation+ integrated focal element filtering*. The approximated curves well represent the true cumulative Belief curve and correctly identify the steep drop with an error of maximum 0.1 in the Belief values. Plausibility is underestimated but the approximation technique was geared towards an accurate estimation of the Belief. Therefore, an underestimation of the Plausibility was expected. The reduction in computational cost however, is substantial reaching over 80% (see Bel90 with filter 0.1, that corresponds to trust factor of 0.90 and a filter accuracy of 0.1). Figure 7 shows the equivalent result for problem

MV8. In this case the error in Belief for a trust factor of 0.90 is larger than 0.1 but for a very small difference in ν. The overall error in the decision of the reliability the estimation of a design margin would be contained but with a massive reduction in computational cost, up to 85%. It is also interesting to note that for all test cases the solution of the min/max problem returned the global maximum and an optimal **d**. IDEA was set with a population of 10 individuals for the inner loop and 500$n/2$ number of function evaluations. The population was restarted when the maximum distance among individuals was reaching 10% of the maximum distance experienced during the whole search. vIDEA was set with a population of 10 individuals and 5000$n/2$ function evaluations. The population was restarted when the maximum distance among individuals was reaching 10% of the maximum distance experienced during the whole search, and the probability of running the global search in the inner loop was $p_d$=0.5.

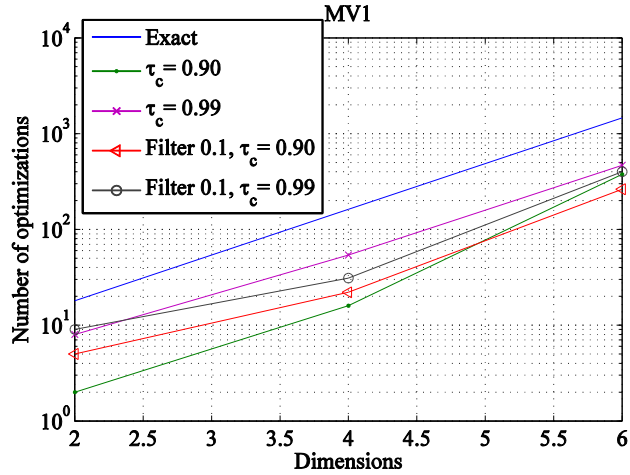

Figure 3. MV1: complexity reduction

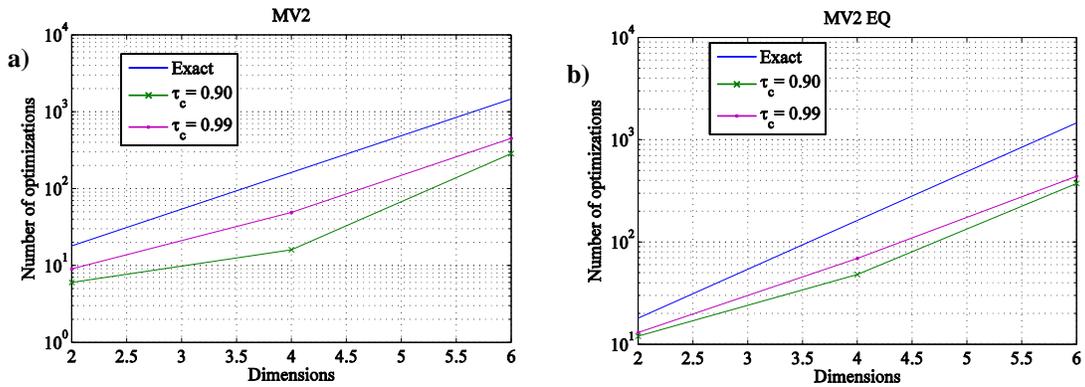

Figure 4 MV2: complexity reduction a) non uniform bpa structure b) uniform bpa structure

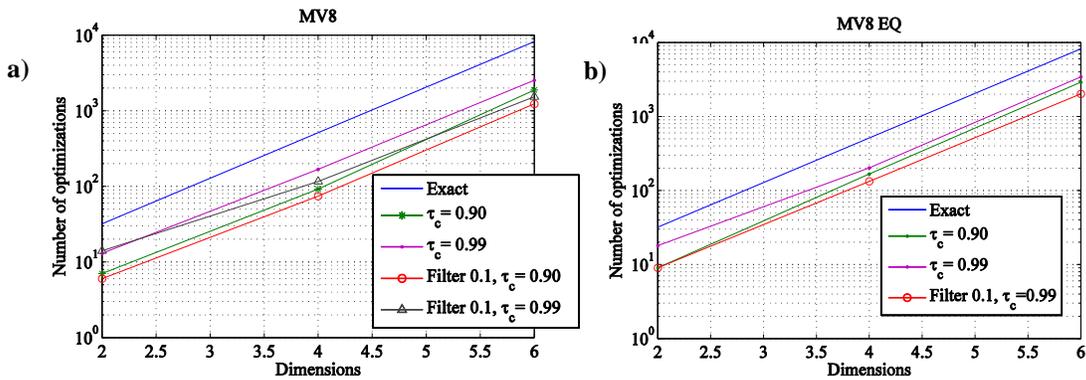

Figure 5 MV8: complexity reduction a) non uniform bpa structure b) uniform bpa structure

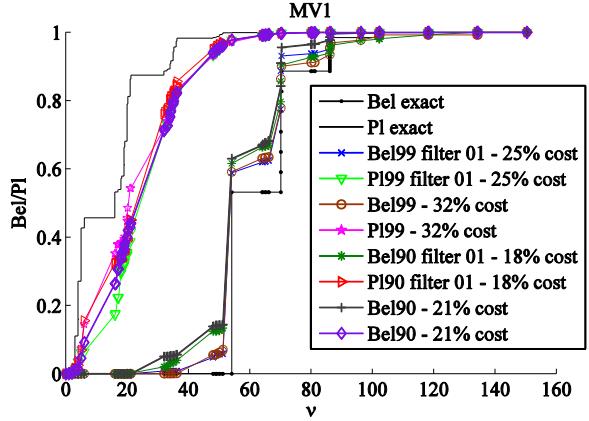

**Figure 6 MV1: Bel and Pl curves at different approximation levels**

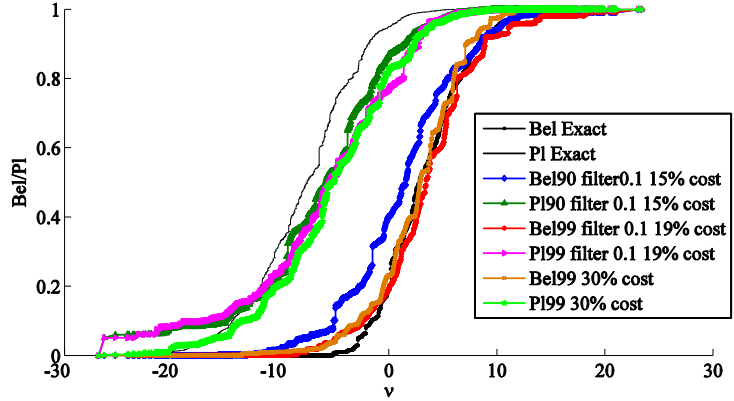

**Figure 7 MV8: Bel and Pl curves at different approximation levels**

### C. POWER-TELECOM INTEGRATED DESIGN PROBLEM

The techniques proposed in this paper were applied to the solution of a realistic case in which an integrated space system made of a power generation unit and telecom subsystem need to be designed under uncertainty. This section describes the power and telecom models used in this paper. The tests in this section aim to show how to use Evidence-Based Robust Optimization can provide a more precise quantification of the design margins, compared to a more traditional approach using rule of the thumb margins. The models in this section are derived from Ref. 16,17,18 and 19.

*3. Power System Model*

The power system (POW) model consists of a solar arrays and a battery. Starting from the required power in daylight and eclipse, the total required power is computed as:

$$P_{sa} = \frac{\left(\frac{P_e T_e}{X_e}\right) + \left(\frac{P_d T_d}{X_d}\right)}{T_d} \qquad (10)$$

where $P_e$ is the power consumption during eclipse, $T_e$ is the orbital eclipse time, $X_e$ is the energy transfer efficiency during eclipse, $P_d$ is the power consumption during daylight, $T_d$ is the orbital daylight time, $X_d$ is the energy transfer efficiency during daylight. The generated power at Beginning of Life (*BOL*)is:

$$P_o = \eta_{cell} G_S \qquad (11)$$

$$P_{BOL} = P_o I_d \cos \alpha_{SA} \qquad (12)$$

where $P_o$ is the ideal power output per unit area of the solar arrays, $\eta_{cell}$ the solar cell efficiency, $G_S$ is the solar flux, $I_d$ is the inherent degradation and $\alpha_{SA}$ is the worst case angle of incidence of the Sun light. In order for the model to calculate the End of Life (*EOL*) power output per unit area, a solar array degradation over satellite lifetime factor $L_d$ is calculated as follows:

$$L_d = (1 - \delta_{cell})^{Life} \tag{13}$$

where $\delta_{cell}$ is the array degradation per year, *Life* is the expected satellite lifetime. Once the satellite lifetime factor $L_d$ is computed, the power output during *EOL*, $P_{EOL}$, can be calculated, based on the power output per unit area $P_{BOL}$, as follows:

$$P_{EOL} = P_{BOL} L_d \tag{14}$$

Then the required solar array area $A_{sa}$ can be easily calculated as:

$$A_{sa} = \frac{P_{sa}}{P_{EOL}} \tag{15}$$

The solar array mass $M_{sa}$ is then derived from the solar array area $A_{sa}$ as follows:

$$M_{sa} = A_{sa} \rho_{sa} \tag{16}$$

where $\rho_{SA}$ is the specific mass of the panel. The cell efficiency $\eta_{cell}$ defines the type of solar cell that will be used including its intrinsic characteristics. For every value of $\eta_{cell}$ a database of cells, see Table 3, is used to obtain the rest of the cell characteristics.

**Table 3. Solar cell intrinsic characteristics.**

|  | CdTe | p c-Si | u c-Si | 3j GaAs | Conc. 3j GaAs | Multijunc. cells |
|---|---|---|---|---|---|---|
| $\eta_{cell}$ | 0.165 | 0.203 | 0.25 | 0.30 | 0.38 | 0.41 |
| $\delta_{cell}$ | 1 | 0.037 | 0.037 | 0.05 | 0.05 | 0.05 |

The PCU power output $P_{pcu}$ is calculated as follows:

$$P_{pcu} = P_{sa} / \eta_{pcu} \tag{17}$$

where $\eta_{pcu}$ is the PCU efficiency. Finally the PCU mass $M_{pcu}$ is calculated as a fraction of the PCU power output:

$$M_{pcu} = a_{pcu} P_{pcu} \tag{18}$$

where $a_{pcu}$ is a PCU mass coefficient. The battery mass $M_{bat\_pack}$, is computed starting from the energy density $E_d$, which defines the particular battery chemistry to be used (see Table 4). The efficiency depends on the type of battery and therefore on $E_d$. The efficiency $\eta_{batt}$ is computed by linearly interpolating the data in Table 4. Furthermore, using a simple linear relationship in logarithmic scale, the depth of discharge *DOD* is calculated as a function of parameter *q* in Table 4 and the number of cycles $N_{cycles}$. The number of cycles is derived from the orbital characteristics and a fixed input in this analysis.

**Table 4. Battery intrinsic characteristics**

|  | NiCd | NiH$_2$ |
|---|---|---|
| $E_d$ (Wh/kg) | 60 | 75 |
| $\eta_{batt}$ (%) | 85 | 86 |
| q |  145.8 | 176.3 |

The minimum required battery capacity $C_{min}$ can then be calculated as follows:

$$C_{min} = \frac{P_e T_e}{DOD \eta_b} \tag{19}$$

and the mass of the battery cells $M_b$ is calculated as:

$$M_b = \frac{C_{min}}{E_d} \tag{20}$$

*4. Telecom System Model*

The mass and power of the telecom system (TTC) are computed starting from the link budget. The required communication link characteristics are the Bit Error Rate *BER*, the modulation, and ground station antenna gain $G_r$. From the *BER* and modulation, one can compute the required energy per bit to noise ratio $E_bN_o$. The $E_bN_o$ plus the data rate are used to compute the Carrier to Noise Ratio $CN_{ratio}$. The total amount of data to be transmitted is assumed to be $T_{data} = 10^3 B$ where *B* is the total amount of data coming from the C&DH (Command & Data Handling) system to telecom. Given the access time $\Delta t_a$ the required data rate $R_t$ is calculated as follows:

$$R_t = 10 \log_{10}\left(\frac{T_{data}}{\Delta t_a - \Delta t_{aq}}\right) \tag{21}$$

where $\Delta t_{aq}$ is the target acquisition time. Given the data rate and the bit to noise ratio, $CN_{ratio}$ is simply:

$$CN_{ratio} = E_b N_0 + R_t \tag{22}$$

With the Carrier to Noise Ratio one can compute the Equivalent Isotropic Radiated Power (EIRP) as follows:

$$EIRP = CN_{ratio} - G/T + L_{TOTAL} - k \tag{23}$$

where *k* = 228.6 dB, $L_{TOTAL}$ is the total signal loss and *G/T* is receiving system performance. The total signal loss is computed adding up all the factors that lead to a loss of signal energy and an increase of the noise. Here most of these losses or sources of noise have been modeled with simple equations or look-up tables. The free space losses $F_{SL}$ are calculated from the distance from the ground station $r_{GS}$ as well as the frequency of the transmitter $f_T$:

$$F_{SL} = 32.4 + 20 \log_{10} r_{GS} + 20 \log_{10} f_T \tag{24}$$

The polarization mismatch (Ionospheric) losses $P_L$ can be computed from the Faraday rotation $\varphi_f$ using the following relationship:

$$P_L = -20 \log_{10}(\cos \varphi_f) \tag{25}$$

The atmospheric losses $A_L$ are a function of the ground station altitude $H_G$, are collected in a look-up table (as in Table 5) and interpolated. The dependency of the atmospheric losses on the elevation angle is modeled by introducing a simple sinusoidal function of the elevation angle *e*:

$$A_{LH} = \frac{A_L}{\sin e} \tag{26}$$

**Table 5. Atmospheric losses' change with ground station altitude**

| $H_G$ (km) | $A_L$ (dB) |
|---|---|
| -2 to 2 | 0.04 |
| 2.1 to 6 | 0.025 |
| 6.1 to 10 | 0.008 |
| 10.1 to 14 | 0.004 |
| 14.1 to 18 | 0.001 |

The Rain absorption losses $Ra_L$ are then calculated by using the data in Ref 16 and 18. The worst case losses for the Feeder loss $F_L$, the Antenna misalignment loss $AM_L$ and the implementation loss $I_L$ are reported in Table 6.

**Table 6. Worst case losses**

| $F_L$ [dB] | $AM_L$ [dB] | $I_L$ [dB] |
|---|---|---|
| 2 | 0.5 | 2 |

Summing up all the individual losses provides the total loss $L_{TOTAL}$:

$$L_{TOTAL} = FS_L + F_L + AM_L + A_{LH} + P_L + R_{aL} + I_L \tag{27}$$

The system noise is computed from the antenna noise temperature $AN_{temp}$ and from the cabling and receiver losses. The total noise gives the noise figure $RN_{fig}$:

$$RN_{fig} = AN_{temp} + T_{AMP} + \frac{\left(10^{L_A/10} - 1\right)k_o + 10^{L_A/10}\left(10^{F/10} - 1\right)k_o}{10^{G_{AMP}/10}} \tag{28}$$

where $T_{AMP}$ is the amplifier noise, $L_A$ the cable loss, $G_{AMP}$ the low noise amplifier gain, *F* the receiver noise figure and $k_0 = 290$. The transmitter noise temperature $S_{temp}$ is:

$$S_{temp} = AT_{tempT} + T_{eT} + \frac{\left(10^{L_T/10} - 1\right)k_o + 10^{L_T/10}\left(10^{F_T/10} - 1\right)k_o}{10^{G_T/10}} \quad (29)$$

Here $AT_{tempT}$ is the transmitter antenna noise temperature, $T_{eT}$ is the transmitter amplifier noise, $L_T$ is the transmitter cable loss, $G_T$ is the transmitter low noise amplifier gain, $F_T$ is the transmitter noise figure. The rain noise $N_{rain}$ is then calculated as follows:

$$N_{rain} = \left(1 - \frac{1}{10^{RA/10}}\right)k_o \quad (30)$$

where $RA$ is the rain absorption. The total system noise $TS_{noise}$ then writes:

$$TS_{noise} = 10\log_{10}\left(RN_{fig} + S_{temp} + N_{rain}\right) \quad (31)$$

The receiving system performance $G/T$ is then calculated as follows:

$$G/T = G_r - TS_{noise} \quad (32)$$

where $G_r$ is the ground station receiver gain. The required transmission power $P_{ld}$ onboard the spacecraft is defined as:

$$P_{Ld} = EIRP - G_t \quad (33)$$

where $G_t$ is the transmitter antenna gain. The spacecraft antenna type is chosen on the basis of the required antenna gain $G_t$. It is well know that the best antenna for 5 dB ≤ gains ≤ 10 dB is the patch one, while the best for 10 dB < gains ≤ 20 dB belongs to the horn type set, therefore the mass of the antenna is computed as follows. The antenna characteristic length (it is the diameter of the normal conical section for conical horns, parabolas, and circular patches, and an equivalent diameter for pyramidal horns and square/rectangular patches) is:

$$D_{ant} = \left(\frac{10^{\frac{G_t}{10}}}{\eta_{ANT}}\right)^{0.5} \frac{c}{\pi f_T} \quad (34)$$

where $\eta_{ANT}$ is the antenna efficiency and c is the speed of light. If $5 \leq G_t \leq 10$dB the mass of the patch is:

$$M_{ant,patch} = \pi \frac{D_{ant}^2}{4}\left(0.0015\,\rho_{diel} + 0.0005\,\rho_{copper}\right) \quad (35)$$

where $\rho_{diel}$ = 2000 kg/m3 and $\rho_{copper}$ = 8940 kg/m3 are the averaged value of a dielectric material density and the copper density, respectively, considering a 2 mm total thickness, with 1.5 mm of dielectric material and 0.5 mm copper. If 10 dB < $G_t$ ≤ 20 dB the lateral surface of the horn, $S_{LAT}$, is computed as a conical surface:

$$S_{LAT} = \pi \frac{D_{ant}}{2}\sqrt{\frac{D_{ant}^2}{4} + L_{horn}^2} \quad (36)$$

and the mass, $M_{ant,horn}$, is:

$$M_{ant,horn} = S_{LAT}\,\rho_A \quad (37)$$

where $L_{horn}$ is the length of the horn antenna can be assumed equal to $2D_{ant}$ from available data, and $\rho_A$ is the areal density, which has a mean value of approximately 15 kg/m² (from available data18). If the gain of the antenna is > 20 dB, the parabola antenna is selected, the diameter of the antenna is computed with Eq.(34), and the mass of the antenna, $M_{ant,par}$, is:

$$M_{ant,par} = \pi \frac{D_{ant}^2}{4}\,\rho_A \quad (38)$$

where the surface density has a typical value of 10 kg/m².

The mass of the amplifier $M_{amp}$ is a function of $P_{Ld}$ (see Ref. 17) as well as the mass of the case $M_{case}$. An identification parameter $T \in [0, 1]$ is used to identify the type of amplifier such that for TWTA type, $T = 0$ and for solid state type $T = 1$. Finally, the casing mass $M_{case}$ is computed as a fraction of the amplifier mass:

$$M_{case} = M_{amp}\rho_{CMR} \quad (39)$$

where $\rho_{CMR}$ is the ratio between the mass of the case and the amplifier mass.

## 5. Test Results

The *bpa* structure and the design space for both the TTC and POW system are summarized in Table 7 and Table 8. The assumption for the integrated system is that the power demand for the TTC, $P_{Ld}$, is added to a fixed power demand of 900W in daylight and 400W in eclipse. In these tests, it is assumed that the spacecraft spends half of the time in eclipse and half in daylight with a maximum solar aspect angle of 15degrees.

Figure 8 and Figure 9 show the *Bel* and *Pl* curves for the TTC system and a comparison to the margin quantification using a traditional margin approach. The cost function *f* is the system mass, i.e. the mass of TTC. Note that some intervals are overlapping. This is an interesting feature of Evidence Theory that allows one to deal with what can be considered as the degree of ignorance on the *bpa* assignment. The assumption is that the spacecraft is operating at 1.5e6 km from the Earth and has an access time of 1000s to a ground station with a receiving antenna with a gain of 60dB. The volume of data is 120000 bits. The lifetime of the mission is assumed to be 4 years. The Faraday rotation is assumed to be 9 degrees, the gain of the ground station antenna 60dB and the BER is 1e-6. The ground station is assumed to be at altitude 0m with the spacecraft at 30 degrees of elevation angle. The gain of the amplifier is 60dB with cable losses of 8dB, a noise temperature of 400K, and a noise figure of 10. The transmitter amplifier gain is assumed to be 20dB with noise temperature of 400K and noise figure of 10. Note that the characteristics of the POW and TTC subsystems were not selected to reflect a real mission scenario but only to test the proposed methodology. With these values, the difference between the optimal and robust solution is about 1kg for the TTC.

**Table 7. TTC bpa structure**

| | | | | | |
|---|---|---|---|---|---|
| $\eta_{ANT}$ | Interval | [0.5 0.6] | [0.65 0.75] | [0.6 0.8] | [0.8 0.95] |
| | bpa | 0.2 | 0.5 | 0.2 | 0.1 |
| $\rho_{CMR}$ | Interval | [0.1 0.2] | [0.25 0.3] | [0.1 0.3] | |
| | bpa | 0.5 | 0.35 | 0.15 | |
| $Lt$ | Interval | [1 2] | [2 3] | [3 5] | |
| | bpa | 0.2 | 0.3 | 0.5 | |
| $T_{ant}$ | Interval | [200 250] | [300 370] | [400 500] | |
| | bpa | 0.1 | 0.6 | 0.3 | |

**Table 8. POW bpa structure**

| | | | | |
|---|---|---|---|---|
| $X_e$ | Interval | [0.5 0.6] | [0.65 0.7] | [0.72 0.75] |
| | Bba | 0.1 | 0.6 | 0.3 |
| $X_d$ | Interval | [0.65 0.7] | [0.75 0.8] | [0.8 0.85] |
| | Bba | 0.2 | 0.6 | 0.2 |
| $I_d$ | Interval | [0.8 0.81] | [0.82 0.83] | [0.83 0.9] |
| | Bba | 0.7 | 0.2 | 0.1 |
| $\eta_{PCU}$ | Interval | [0.5 0.6] | [0.65 0.7] | [0.8 0.9] |
| | Bba | 0.1 | 0.6 | 0.3 |

**Table 9. Design space for TTC and POW**

| Parameter | Low bound | Upper bound |
|---|---|---|
| $f_T$ (MHz) | 7e3 | 11e3 |
| *Mod* | 0 | 1 |
| *T* | 0 | 1 |
| $G_t$ (dB) | 5 | 20 |
| $\eta_{cell}$ | 0.1 | 0.3 |
| $\rho_{sa}$ (kg/m$^2$) | 1 | 2 |
| $a_{PCU}$ (kg/W) | 0.01 | 0.02 |
| $E_d$ (Wh/kg) | 60 | 100 |

The *Bel* margin curve in Figure 8 was generated assuming that a designer is taking the min/min solution (best absolute performance) from problem (8) and adding a 25% margin to the required TTC power and to the mass of the casing of the electronics. Then a system level margin is added to the total mass of the TTC. The system level margin can range from 0% to 25% of the nominal mass of the TTC. For each mass plus system margin the value of Belief and Plausibility was computed (see red and green thick solid lines). In Figure 9 a more conservative choice is made.

A 25% margin is added to the mass of antenna and amplifier and then a system level margin is added as before to the total mass of the TTC. The two figures show that the margin approach either underestimates the Belief or overestimates the margin. Figure 8, in fact, shows that the Belief of the mass corresponding to the maximum system level margin is less than 60%. In the more conservative case, Figure 9, the Belief is 1 but the mass is overestimated by about 0.5kg.

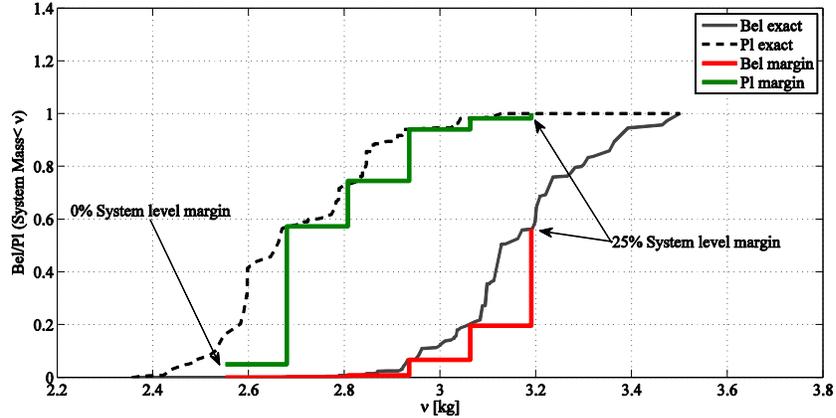

**Figure 8. TTC system : margin approach vs. EBRO: best case margins.**

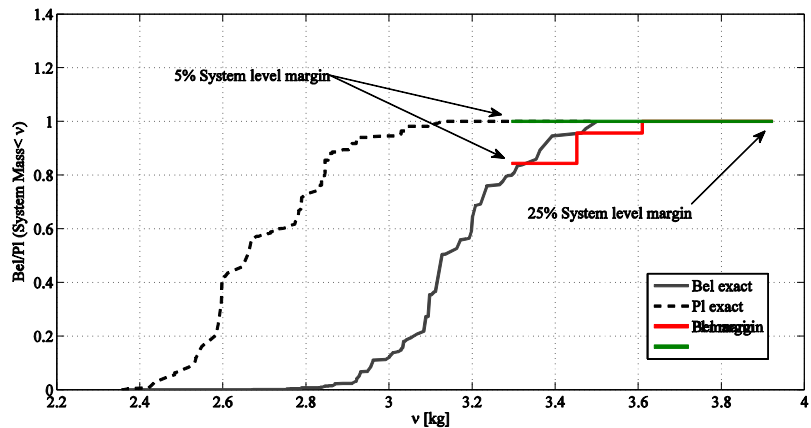

**Figure 9. TTC system: margin approach vs. EBRO: worst case margins.**

Figure 10 shows the *Bel* and *Pl* curves computed with different $\tau_c$ from 0.7 to 0.99. The total number of focal elements is 108 corresponding to 216 optimizations to compute the exact *Bel* and *Pl*. With $\tau_c$= 0.99 one can obtain a reduction of the computational cost down to 26% and with no relevant error in Belief. A $\tau_c$= 0.70 bring a reduction down to 8% of the cost for an exact computation but with a limited error. In particular the error is very contained for high values of Belief with a difference of about 0.1 kg in mass for the same Belief of less than 0.1 difference in Belief for the same mass.

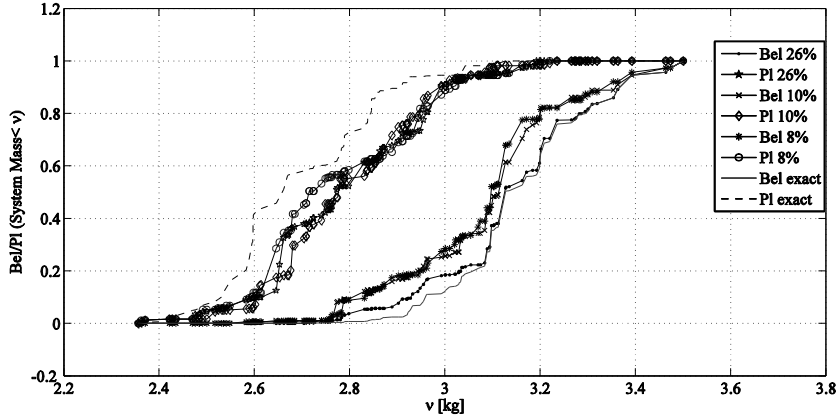

**Figure 10. Approximated *Bel* and *Pl* curves for the Telecom system**

Figure 11 shows a similar result for the integrated Power and Telecom system. The total number of focal elements in this case is 8748 corresponding to 17496 optimizations for an exact calculation. The simple application of the integrated filtering technique beings a moderate reduction of computational effort down to 80% of the cost of the exact computation. The resulting Belief is underestimated by maximum 0.1. The application of the *approximated min and max evaluation* with $\tau_c = 0.90$ brings to a more substantial reduction, down to 22% and with a moderate overestimation that reduces almost to zero close to the left and right extremes of the Belief curve.

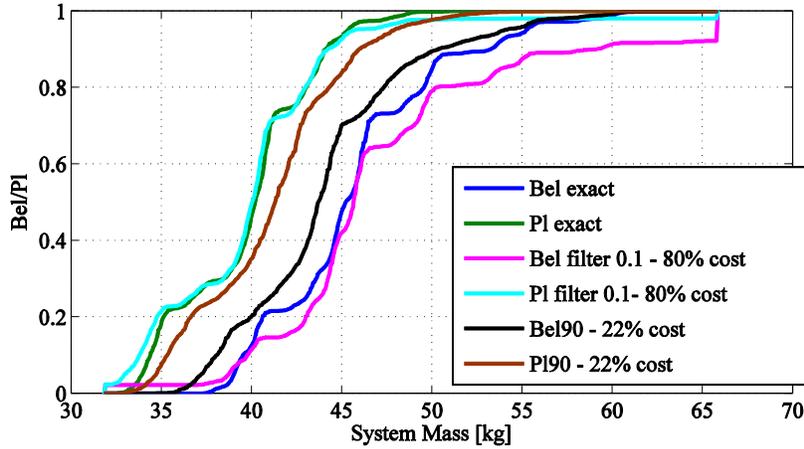

**Figure 11. Approximated *Bel* and *Pl* curves for the integrated Power and Telecom system.**

## VI PROBLEM DECOMPOSITION

The interesting aspect of space engineering systems is that although the overall design requires the contribution of all the subsystems, some subsystems are relatively decoupled and exchange information only through their specific design budgets. For example, the telecom system and the power system exchange information only through the output power from the telecom system that becomes an input parameter to the power system.

Let us consider a function $g : D \times \bar{U} \to \mathbb{R}$ with the following form:

$$g(\mathbf{d}_1,\mathbf{u}_1,...,\mathbf{d}_i,\mathbf{u}_i,\mathbf{u}_{3i},...,\mathbf{d}_{N_f},\mathbf{u}_{N_f},\mathbf{u}_{3N_f}) = f_1(\mathbf{d}_1,\mathbf{u}_1,...,\mathbf{d}_i,\mathbf{u}_i,...,\mathbf{d}_{N_f},\mathbf{u}_{N_f}) + \sum_{i=2}^{N_f} f_i(\mathbf{d}_i,\mathbf{u}_i,\mathbf{u}_{3i}) \quad (40)$$

Now assume that the functional dependency of function $f_1$ on design and uncertain parameters $\mathbf{d}_i$ and $\mathbf{u}_i$ is realized through a function $h_i$ such that:

$$g(\mathbf{d}_1,\mathbf{u}_1,...,\mathbf{d}_i,\mathbf{u}_i,\mathbf{u}_{3i},...,\mathbf{d}_{N_f},\mathbf{u}_{N_f},\mathbf{u}_{3N_f}) = f_1(\mathbf{d}_1,\mathbf{u}_1,...,h_i(\mathbf{d}_i,\mathbf{u}_i),...,h_{N_f}(\mathbf{d}_{N_f},\mathbf{u}_{N_f})) + \sum_{i=2}^{N_f} f_i(h_i(\mathbf{d}_i,\mathbf{u}_i),\mathbf{u}_{3i}) \quad (41)$$

If $h_i$ could be handled as independent variable the two functions $f_1$ and $f_i$ could be decoupled and $Bel(g<v)$ could be expressed as:

$$Bel_d(g \leq v) = \sum_{\substack{\forall \theta_1 \subseteq A_1 \\ \forall \theta_i \subseteq A_i}} \prod_i^{N_f} m(\theta_1) m(\theta_i)$$

$$A_1 = \{\mathbf{u}_1 \in \bar{U}_1 \subseteq \bar{U}, h \mid g \leq v\}$$

$$A_i = \{h_i, \mathbf{u}_{3i} \in \bar{U}_{3i} \subseteq \bar{U} \mid g \leq v\} \quad (42)$$

If the value of the design parameters and of $h_i$ is fixed then the two functions are completely decoupled and the computation of the Belief associated to their sum requires the independent computation of the maxima and minima of $f_1$ and $f_i$ over the subspaces $\bar{U}_1$ and $\bar{U}_{3i}$. If the range of $h_i$ is well defined then one could compute the values of $f_i$ for different values of $h_i$ by solving the following constrained problems:

$$\max_{\mathbf{u}_{3i}} f_i(\bar{\mathbf{d}}, \mathbf{u}_{3i}, h_i)$$

$$h_i(\bar{\mathbf{d}}, \mathbf{u}_i) = v_{hi} \quad (43)$$

If $h_i$ is fixed the computational complexity grows linearly $N_f$ and the computation of the focal elements for each function $f_i$ can be performed in parallel. If the vector $\mathbf{u}_i$ contains a single uncertain parameter the result is an exact representation of the Belief curve. If the vector $\mathbf{u}_i$ contains multiple uncertain parameters then one can verify that the belief function $Bel_d$ is equal to the full belief curve $Bel$ only for the $v$ values for which the value of $h_i$ is verified. A good choice to fix the value of $h_i$ is to take the solution of the min-max problem (7). As an illustrative example consider the following toy problem:

$$f_2 = d_1^2 + d_2^2 - u_{21}^2 - u_{22}^2 + 2 + u_{32}$$

$$h = u_{21} + u_{22}$$

$$f_1 = d_3^2 + u_1 + h \quad (44)$$

$$g = f_1 + f_2$$

Each uncertain parameter is defined over the three intervals [0 0.1], [0.2 0.4] and [0.5 1] with two different *bpa* assignments: a) 0.3, 0.6 and 0.1 respectively and b) 0.3, 0.1 and 0.6 respectively. Figure 12 shows a comparison between the exact *Bel* and *Pl* curves and the approximated ones. The approximated *Bel* curves are very close to the exact ones and are almost identical for some values of ν. The *Pl* approximation is instead very poor for low values of ν and good for high values of ν. The main reason is that only one value of *h* as used and it was the one corresponding to the solution of the min/max problem. All minimum values within the focal elements of the decomposed problem are therefore overestimated while the maxima are exact for a large number of focal elements at some specific ν. The main advantage of this approximation becomes clear when one looks at the computational cost. The exact computation requires 162 optimizations, while the approximated computation requires two parallel sets of optimizations with 3 optimizations per set, thus 6 in total. The computational cost of the approximation is therefore 3.7% of the cost of the exact computation and would grow linearly with the number of dimensions.

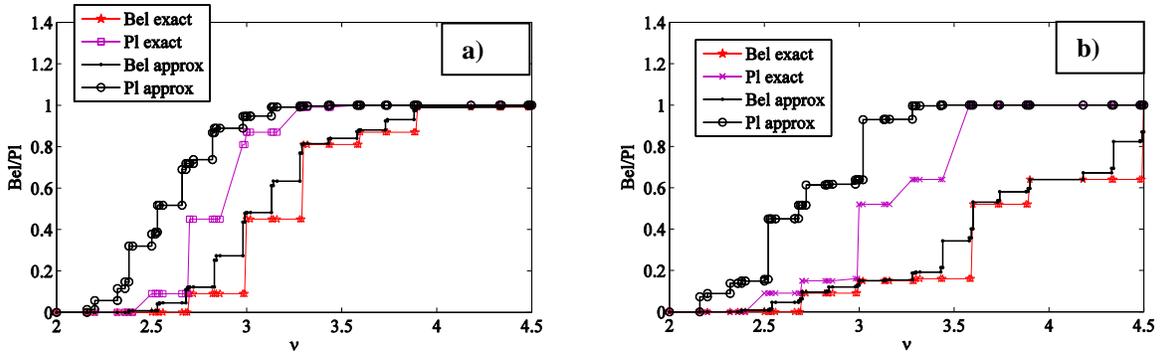

**Figure 12. Decomposition approximation for *bpa* structure a) and *bpa* structure b)**

## VII   FINAL REMARKS

The paper presented some strategies to obtain an estimation of Belief and Plausibility at a fraction of the computational cost for their exact calculation. The approach presented in this paper provides the computation of the optimal range of the design margin at a cost that is polynomial with the number of dimensions. An estimation of the full Belief and Plausibility corves could be obtained with a cost reduction by over 80% but maintaining a contained error. The effectiveness of the proposed strategies was proven on some benchmark problems, presenting a number of minima and maxima exponentially increasing with the number of dimensions. Furthermore, two space system design cases are used to show how evidence-based design optimization can improve the design of space systems compared to a more traditional system margin approach. From these test it appeared that, even in the ideal case in which an optimal deterministic design solution is available, a traditional margin approach tend to underestimate the reliability of the design margin or to overestimate their value, given the available information. This justifies the use of a rigorous margin quantification. Finally, a problem decomposition technique was proposed to reduce the computational complexity of space system design problems in which all the components contributing to the overall design budgets are only weakly coupled through a single function (the power in the case of space systems). For these particular problems, it seems possible to obtain massive reductions of the computational cost but, more importantly, a computational cost that increases linearly with the number of integrated systems. The results in this paper, however, are only preliminary and a more in depth investigation is underway.

## Acknowledgments

This work is partially supported through an ESA/ITI grant AO/1-5679/08/NL/CB
.